\documentstyle[12pt,axodraw]{article}
\newcommand{\EPJ}{Eur. Phys. J. }
\newcommand{\IJMP}{Int. J. Mod. Phys. }
\newcommand{\JHEP}{J. High Energy Phys. }
\newcommand{\NP}{Nucl. Phys. }
\newcommand{\PR}{Phys. Rev. }

\newcommand{\PL}{Phys. Lett. }

\newcommand{\tildep}{\tilde{p}}
\newcommand{\tildek}{\tilde{k}}

\begin{document}
\baselineskip=20pt

\pagenumbering{arabic}

\vspace{1.0cm}
\begin{flushright}
LU-ITP 2002/022 
\end{flushright}

\begin{center}
{\Large\sf Spectral representation and dispersion relations in field theory 
on noncommutative space}\\[10pt]
\vspace{.5 cm}

{Yi Liao, Klaus Sibold}
\vspace{1.0ex}

{\small Institut f\"ur Theoretische Physik, Universit\"at Leipzig,
\\
Augustusplatz 10/11, D-04109 Leipzig, Germany\\}

\vspace{2.0ex}

{\bf Abstract}

\end{center}

We study the spectral representation and dispersion relations that follow from 
some basic assumptions and the reduced spacetime symmetries on noncommutative 
(NC) space. Kinematic variables involving the NC parameter appear naturally as 
parametric variables in this analysis. When subtractions are necessary to 
remove ultraviolet divergences, they are always made at the fixed values of 
these NC variables. This point is also illustrated by a perturbative analysis 
of self-energies. Our analysis of the reduced spacetime symmetries suggests a 
weaker microcausality requirement. Starting from it, we make a first attempt at 
dispersion relations for forward scattering. It turns out that the attempt is 
hampered by a new unphysical region specified by a given motion in the NC 
plane which does not seem to be surmountable using the usual tricks. 
Implications for a possible subtraction and renormalization scheme for NC 
field theory in which the ultraviolet-infrared (UV/IR) mixing is removed 
are also briefly commented on.  

\begin{flushleft}


\end{flushleft}

\newpage
\section{Introduction}

Quantum field theory on noncommutative (NC) spacetime has some intriguing 
features that result from its nonlocal nature of interactions. The quantum 
corrections that are effectively regularized in the ultraviolet (UV) by 
the nonlocality become singular again for some kinematic configurations 
including the infrared (IR) one. This is the so-called UV/IR mixing problem 
$\cite{mixing}$ 
that hampers the implementation of the usual renormalization procedure. There 
are additional problems when time does not commute with space. A direct 
application of Feynman rules for which the only NC effect is 
the appearance of NC phases $\cite{filk}$ leads to violation of unitarity 
$\cite{unitarity}$. However, as has been shown recently, a proper treatment of 
perturbation theory in this case does not yield the assumed Feynman rules 
and unitarity can be maintained without any problem $\cite{bahns,liao}$. 
Another issue with time-space NC is that it seems to lead to acausal effects 
in scattering $\cite{causality}$. Although there is no solution to this so far, 
there is a hope that this could be resolved by a proper redefinition of 
wavefunctions and states in the NC context $\cite{lubo}$. 

The above interesting results are based on explicit perturbative analysis. In 
this work we shall study the related issue concerning analytic properties of 
Green functions and scattering amplitudes by starting from some basic 
assumptions in NC field theory. In particular we want to know how the 
K\"allen-Lehmann spectral representation can be generalized to the NC case, and 
whether it is still possible to use microcausality to derive simple dispersion 
relations for scattering amplitudes. 
As we shall see, the drastic changes in locality and spacetime symmetries 
will make things not so obvious and present new obstacles especially in 
deriving a dispersion relation for scattering. Our discussion will also 
give a clue on how to implement the renormalization program in the presence 
of the UV/IR mixing. 

An essential difference of NC field theory from ordinary theory is 
that Lorentz invariance is generally lost. Since the latter is one of the basic 
assumptions in ordinary field theory and facilitates important arguments in 
deriving dispersion relations, we shall first discuss the new situation 
regarding spacetime symmetries due to the introduction of noncommutativity, 
$[\hat{x}_{\mu},\hat{x}_{\nu}]=i\theta_{\mu\nu}$. Our results will be directly 
applicable to NC scalar field theory formulated in terms of the star product. 

First of all, the translational invariance is preserved which guarantees the 
energy-momentum conservation in all processes. We shall be only concerned in 
this work with space-space NC. In this case, without loss of generality, we can 
always choose NC to be restricted in the $(12)$-plane; i.e., 
$\theta_{12}=-\theta_{21}=\theta>0$, with others vanishing. Since 
$\hat{x}_{0,3}$ are not involved in NC, invariance in Lorentz boosts remains 
in the $3$-direction. The NC relation 
$[\hat{x}_1,\hat{x}_2]=i\theta$ is also invariant under 
rotations around the axis in the $3$-direction. Suppose the NC vector 
$\vec{\theta}$ is in the normal direction to the NC plane, then the remaining 
symmetries are summarized as boosts along and rotations around 
$\vec{\theta}$. In momentum space, typical examples of invariants involving 
momenta $p,q$ and $\theta_{\mu\nu}$ include 
$p\cdot\tilde{q}$, $\tilde{p}^2$ and $\tilde{p}\cdot\tilde{q}$, 
where $\tilde{p}_{\mu}=\theta_{\mu\nu}p^{\nu}$ is the NC momentum orthogonal to 
$p$. Denoting the components of the spatial momentum $\vec{p}$ in the 
$\vec{\theta}$ direction and in the NC plane as $p_{\parallel}$ and 
$\vec{p}_{\bot}$ respectively, we have, 
$p\cdot\tilde{q}=\vec{\theta}\cdot(\vec{p}_{\bot}\times\vec{q}_{\bot})$, 
$\tilde{p}^2=-\theta^2\vec{p}_{\bot}^{~2}$, and 
$\tilde{p}\cdot\tilde{q}=-\theta^2\vec{p}_{\bot}\cdot\vec{q}_{\bot}$.
These invariants are in addition to the ordinary ones $p^2$ and $p\cdot q$.

For completeness, we also mention briefly the other cases concerning 
$\theta_{\mu\nu}$. For $\theta_{ij}=0$, we may choose 
$\theta_{03}=-\theta_{30}=\theta$, with others vanishing. The remaining 
symmetries are precisely the same as in the above case. The mixed case of 
$\theta_{0i}\neq 0$ and $\theta_{ij}\neq 0$ is generally complicated and there 
are usually no remaining symmetries separately for boosts and rotations except 
for one special case. Defining 
$e^i=\theta^{0i}$ and $b^i=1/2~\epsilon^{ijk}\theta^{jk}$, the special 
case corresponds to $\vec{e}\parallel\vec{b}$ for which invariance remains 
in boosts along and rotations around the preferred direction.

The rest of the paper is organized as follows. We study in the next section the 
two-point function and its spectral representation. A dispersion relation is 
then written down using the analytic properties implied in the representation. 
As an example, we demonstrate explicitly in section $3$ that the dispersion 
relation is satisfied for the self-energy at one loop in $\varphi^3$ 
perturbation theory. 
In section $4$ we attempt to derive a dispersion relation for forward 
scattering amplitudes. To this end we propose a microcausality requirement 
which is weaker than the usual one but is in accord with our symmetry analysis 
in this section. Its viability is further exemplified by a perturbative 
calculation. We show how a new unphysical region arises that hampers any 
simple dispersion relations for scattering amplitudes. 
We summarize our results and discuss their implications in the last section.

\section{Spectral representation} 

We assume that the usual assumption in ordinary field theory about physical 
spectrum and its completeness is still a reasonable starting point in 
space-space NC field theory. The main difference is that only part 
of relativistic invariance is preserved in NC theory. Our discussion follows 
the usual development in axiomatic field theory $\cite{dispersion}$ 
and is a generalization of the 
K\"allen-Lehmann spectral representation to the NC space.

Let us consider the two-point function defined in terms of the Heisenberg 
scalar field $\varphi$ and vacuum state $|\Omega\rangle$, 
\begin{equation}
\begin{array}{rcl}
D^{\prime}(x,y)&=&\langle\Omega|\varphi(x)\varphi(y)|\Omega\rangle\\
&=&D^{\prime}(x-y), 
\end{array}
\end{equation}
where translational invariance has been used. Using the completeness relation 
of the physical spectrum, we have, 
\begin{equation}
\begin{array}{rcl}
D^{\prime}(z)&=&\displaystyle
\sum_{\alpha}\int\frac{d^3\vec{p}}{(2\pi)^32E_{\vec{p},\alpha}}
\left|\langle\Omega|\varphi(0)|\vec{p},\alpha\rangle\right|^2e^{-ip_+\cdot z}\\ 
&=&\displaystyle 
\sum_{\alpha}\int \frac{d^4p}{(2\pi)^3}\tau(p_0)\delta(p^2-m^2_{\alpha})
\left|\langle\Omega|\varphi(0)|\vec{p},\alpha\rangle\right|^2e^{-ip\cdot z}. 
\end{array}
\end{equation}
Here $z=x-y$, $p_+^{\mu}=(E_{\vec{p},\alpha},\vec{p})$ and 
$E_{\vec{p},\alpha}=\sqrt{\vec{p}^{~2}+m^2_{\alpha}}$ with $\alpha$ representing 
all other quantum numbers specifying a state. $\tau$ is the step function; we 
have reserved $\theta$ for the NC parameter.

In ordinary theory the matrix element appearing in the above equation is a 
Lorentz invariant and depends on $p$ only through $p^2=m^2_{\alpha}$. In the NC 
case, the symmetry argument is less restrictive so that additional dependence 
on $p$ is permitted. According to our discussion in the previous section, this 
dependence can occur only in the form of 
$\tilde{p}^2=-\theta^2\vec{p}_{\bot}^{~2}$. 
Inserting the $\delta$ function identity, we obtain 
\begin{equation}
\begin{array}{rcl}
D^{\prime}(z)&=&\displaystyle
\int_0^{\infty} dm^2\sum_{\alpha}\int \frac{d^4p}{(2\pi)^3}
\tau(p_0)\delta(p^2-m^2)\delta(m^2-m^2_{\alpha})
\left|\langle\Omega|\varphi(0)|\vec{p},\alpha\rangle\right|^2e^{-ip\cdot z}\\
&=&\displaystyle 
\int dm^2\int \frac{d^4p}{(2\pi)^3}\tau(p_0)\delta(p^2-m^2)
e^{-ip\cdot z}\rho(m^2,\tildep^2)\\ 
&=&\displaystyle 
\int dm^2\rho(m^2,(i\tilde{\partial})^2)
\int \frac{d^4p}{(2\pi)^3}\tau(p_0)\delta(p^2-m^2)e^{-ip\cdot z}\\ 
&=&\displaystyle 
\int dm^2\rho(m^2,(i\tilde{\partial})^2)D(z,m^2), 
\end{array}
\end{equation}
where
$(\tilde{\partial})^2=\theta_{\mu\beta}\theta^{\mu}_{~\gamma}
\partial^{\beta}_z\partial^{\gamma}_z
=-\theta^2[(\partial_{z^1})^2+(\partial_{z^2})^2]$, 
$\rho(m^2,\tildep^2)$ 
is the spectral density function for a fixed $\tildep^2$, 
\begin{equation}
\begin{array}{rcl}
\rho(m^2,\tildep^2)&=&\displaystyle 
\sum_{\alpha}\delta(m^2-m^2_{\alpha})
\left|\langle\Omega|\varphi(0)|\vec{p},\alpha\rangle\right|^2, 
\end{array}
\end{equation}
and $D$ is the usual function defined for free fields, 
\begin{equation}
\begin{array}{rcl}
D(z,m^2)&=&\displaystyle 
\int \frac{d^4p}{(2\pi)^3}\tau(p_0)\delta(p^2-m^2)e^{-ip\cdot z}. 
\end{array}
\end{equation}

The above representation has interesting physical implications. Let us first 
consider the following vacuum expectation value of the commutator, 
\begin{equation}
\begin{array}{rcl}
i\Delta^{\prime}(z)&=&\langle\Omega|[\varphi(x),\varphi(y)]|\Omega\rangle\\
&=&\displaystyle 
\int dm^2\rho(m^2,(i\tilde{\partial})^2)i\Delta(z,m^2), 
\end{array}
\end{equation}
with $i\Delta(z,m^2)=D(z,m^2)-D(-z,m^2)$. It is well-known that the 
distribution $\Delta(z,m^2)$ is singular on the light-cone and vanishes for 
space-like $z$. The subtle point here is that $\rho(m^2,\tildep^2)$ is 
generally not a polynomial in $\tildep^2$ of finite order. The integrand in the 
above equation involves derivatives of infinite order and is thus highly 
nonlocal in spatial $z$. It is not clear whether it persists to vanish for 
space-like $z$ after this manipulation of derivatives. A more careful analysis 
will be given in section $4$. The result is that $\Delta^{\prime}(z)$ vanishes 
when $z$ is space-like in the commutative direction, i.e., 
$z_0^2<z_{\parallel}^2$. 

The second quantity that we would like to explore is the complete Feynman 
propagator, 
\begin{equation}
\begin{array}{rcl}
iD_F^{\prime}(z)&=&\displaystyle 
\langle\Omega|T(\varphi(x)\varphi(y))|\Omega\rangle\\
&=&\displaystyle \tau(z_0)D^{\prime}(z)+\tau(-z_0)D^{\prime}(-z)\\
&=&\displaystyle \int_0^{\infty}dm^2\rho(m^2,(i\tilde{\partial})^2)
[\tau(z_0)D(z,m^2)+\tau(-z_0)D(-z,m^2)]\\
&=&\displaystyle \int dm^2\rho(m^2,(i\tilde{\partial})^2)
iD_F(z,m^2), 
\end{array}
\end{equation}
where $iD_F$ is the Feynman propagator for free fields, 
\begin{equation}
\begin{array}{rcl}
iD_F(z,m^2)&=&\displaystyle 
\int\frac{d^4p}{(2\pi)^4}\frac{i}{p^2-m^2+i\epsilon}e^{-ip\cdot z}.
\end{array}
\end{equation}
We stress that the third equality is possible only for $\theta_{0i}=0$; 
otherwise the derivative operation cannot commute with the step function. 
Transforming into momentum space yields, 
\begin{equation}
\begin{array}{rcl}
i\hat{D}_F^{\prime}(k)&=&\displaystyle 
\int d^4z e^{ik\cdot z}iD_F^{\prime}(z)\\
&=&\displaystyle 
\int_0^{\infty} dm^2\rho(m^2,\tildek^2)\int d^4z e^{ik\cdot z}iD_F(z,m^2)\\
&=&\displaystyle 
\int dm^2\rho(m^2,\tildek^2)i\hat{D}_F(k^2,m^2),
\end{array}
\end{equation}
where $i\hat{D}_F(k^2,m^2)=i(k^2-m^2+i\epsilon)^{-1}$ is the Feynman propagator 
for free fields in momentum space. In deriving the second equality above, we 
have employed integration by parts in $z$ and ignored spatial surface terms. 
This was done in the same spirit as we cope with the cyclicity of star products 
in the action, and is consistent with the energy-momentum conservation which 
implies that total derivative terms are ignorable in the action. The above 
equation thus generalizes the K\"allen-Lehmann representation to the NC space, 
\begin{equation}
\begin{array}{rcl}
\hat{D}_F^{\prime}(k^2,\tildek^2)&=&\displaystyle 
\int_0^{\infty} dm^2\frac{\rho(m^2,\tildek^2)}{k^2-m^2+i\epsilon}.
\end{array}
\end{equation}

The NC momentum $\tildek$ appears here as a parametric variable. 
Thus, the ordinary arguments leading from the above representation to 
a dispersion relation still apply; namely, for independent but fixed 
$\tildek^2$, $\hat{D}_F^{\prime}(k^2,\tildek^2)$ is an analytic function in 
the complex $k^2$ plane except for some simple poles and a branch cut in the 
positive axis due to the occurrence of isolated states and continum thresholds. 
We can therefore write down the following Hilbert transform, 
\begin{equation}
\begin{array}{rcl}
\hat{D}_F^{\prime}(k^2,\tildek^2)&=&\displaystyle 
\frac{1}{\pi}\int_{-\infty}^{\infty} dk^{\prime 2}
\frac{{\rm Im}~\hat{D}_F^{\prime}(k^{\prime 2},\tildek^2)}
{k^{\prime 2}-k^2-i\epsilon}, 
\end{array}
\end{equation}
which is the dispersion relation for the two-point Green function on NC space. 
The lower delimiter in the above equation can be effectively replaced by the 
squared mass of the lowest physical state. In most physical situations where 
the integrand does not converge fast enough as $k^{\prime 2}\to\infty$, 
subtractions are necessary to make the integral well-defined. We stress that 
these subtractions in $k^2$ must be made for the same fixed $\tildek^2$. 

\section{Dispersion relation for self-energy: a perturbative example} 

To illustrate analytic properties of the two-point functions studied in the 
preceding section, we consider here an explicit one-loop example in 
perturbation theory. A similar analysis has been made previously for massless 
fields at the same level by a straightforward but complicated computation of 
loop integrals $\cite{brandt}$. Our method will be simpler without actually 
evaluating loop integrals and more general in that it is applicable to massive 
fields as well. 

The complete Feynman propagator $\hat{D}^{\prime}(k^2,\tildek^2)$ 
is related to the 1PI self-energy $\Sigma(k^2,\tildek^2)$ by, 
\begin{equation}
\begin{array}{rcl}
\hat{D}^{\prime}(k^2,\tildek^2)&=&\displaystyle 
[k^2-m^2-\Sigma(k^2,\tildek^2)+i\epsilon]^{-1}.
\end{array}
\end{equation}
Note that, as the 1PI functions are defined by Legendre transform from the 
complete Green functions, this relation itself does not invoke for perturbation 
theory. $\Sigma$ has similar analytic properties as $\hat{D}^{\prime}$
except that the latter has additional simple poles. We therefore have, for 
sufficently converging $\Sigma$ at infinity, 
\begin{equation}
\begin{array}{rcl}
\Sigma(k^2,\tildek^2)&=&\displaystyle 
\frac{1}{\pi}\int_{-\infty}^{\infty} dk^{\prime 2}
\frac{{\rm Im}~\Sigma(k^{\prime 2},\tildek^2)}
{k^{\prime 2}-k^2-i\epsilon}. 
\end{array}
\end{equation}
For $\Sigma$ going to a constant as $k^2\to\infty$, one subtraction is 
necessary for the same fixed $\tildek^2$, 
\begin{equation}
\begin{array}{rcl}
\Sigma(k^2,\tildek^2)-\Sigma(\mu^2,\tildek^2)&=&\displaystyle 
\frac{k^2-\mu^2}{\pi}\int_{-\infty}^{\infty} dk^{\prime 2}
\frac{{\rm Im}~\Sigma(k^{\prime 2},\tildek^2)}
{(k^{\prime 2}-k^2-i\epsilon)(k^{\prime 2}-\mu^2-i\epsilon)}, 
\end{array}
\label{selfenergy}
\end{equation}
where $\mu^2$ is an arbitrary subtraction point. Now we demonstrate that the 
above is satisfied at one loop in $\varphi^3$ theory.

The self-energy is, in $n$ dimensions and up to factors of the coupling 
constant, 
\begin{equation}
\begin{array}{rcl}
\Sigma(k^2,\tildek^2)&=&\displaystyle 
i\int\frac{d^n\ell}{(2\pi)^n}\frac{\cos(\ell\cdot\tildek)+1}
{(\ell^2-m^2+i\epsilon)((\ell+k)^2-m^2+i\epsilon)}\\ 
&=&\displaystyle 
i\int_0^1dx\int\frac{d^n\ell}{(2\pi)^n} 
\frac{\cos(\ell\cdot\tildek)+1}{[\ell^2+k^2x(1-x)-m^2+i\epsilon]^2}, 
\end{array}
\end{equation}
where $k$ and $\ell$ are the external and loop momentum respectively. 
For $\theta_{0i}=0$, the cosine function does not contain the $\ell_0$ 
component. There is thus no new obstacle compared to ordinary theory in 
continuing $\ell_0$ analytically to its imaginary axis. The analytic properties 
in $k^2$ will still be governed by the denominator while the numerator serves 
to modulate the weight in each direction according to the spatial NC. Using the
subscript $E$ to indicate the Euclidean loop momentum, we have, 
\begin{equation}
\begin{array}{rcl}
\Sigma(k^2,\tildek^2)&=&\displaystyle 
-\int_0^1dx\int\frac{d^n\ell_E}{(2\pi)^n} 
\frac{\cos(\ell_E\cdot\tildek)+1}{[\ell_E^2+m^2-k^2x(1-x)-i\epsilon]^2}\\
&=&\displaystyle 
\frac{d}{dm^2}\int_0^1 dz\int\frac{d^n\ell_E}{(2\pi)^n} 
\frac{\cos(\ell_E\cdot\tildek)+1}{\ell_E^2+m^2-k^2(1-z^2)/4-i\epsilon}, 
\end{array}
\end{equation}
where we have changed the variable $x\to z=1-2x$ and noted that the integrand 
is even in $z$. An imaginary part develops when the denominator can vanish in 
the integrated domain, which is possible only when $k^2\ge 4m^2$, 
\begin{equation}
\begin{array}{rcl}
{\rm Im}~\Sigma(k^2,\tildek^2)&=&\displaystyle 
\pi\frac{d}{dm^2}\int_0^1 dz\int\frac{d^n\ell_E}{(2\pi)^n} 
[\cos(\ell_E\cdot\tildek)+1]\delta[\ell_E^2+m^2-k^2(1-z^2)/4]\\ 
&=&\displaystyle 
\frac{4\pi}{k^2}\frac{d}{dm^2}\int\frac{d^n\ell_E}{(2\pi)^n} 
[\cos(\ell_E\cdot\tildek)+1]\int_0^1 dz\delta(z^2-v^2)\\ 
&=&\displaystyle 
\frac{2\pi}{k^2}\frac{d}{dm^2}\int\frac{d^n\ell_E}{(2\pi)^n} 
[\cos(\ell_E\cdot\tildek)+1]v^{-1}\tau(k^2-4(\ell_E^2+m^2)), 
\end{array}
\end{equation}
with $v=\sqrt{1-4(\ell_E^2+m^2)/k^2}$. 

On the other hand, the once-subtracted self-energy is 
\begin{equation}
\begin{array}{l}
\Sigma(k^2,\tildek^2)-\Sigma(\mu^2,\tildek^2)=\displaystyle 
(k^2-\mu^2)\frac{d}{dm^2}\int_0^1 dz\int\frac{d^n\ell_E}{(2\pi)^n}\\ 
\displaystyle 
\times\frac{[\cos(\ell_E\cdot\tildek)+1](1-z^2)/4}
{[\ell_E^2+m^2-k^2(1-z^2)/4-i\epsilon]
[\ell_E^2+m^2-\mu^2(1-z^2)/4-i\epsilon]}. 
\end{array}
\end{equation}
Changing the variable $z\to k^{\prime 2}=4(\ell_E^2+m^2)/(1-z^2)$ gives, 
\begin{equation}
\begin{array}{rcl}
\Sigma(k^2,\tildek^2)-\Sigma(\mu^2,\tildek^2)&=&\displaystyle 
(k^2-\mu^2)\frac{d}{dm^2}\int\frac{d^n\ell_E}{(2\pi)^n}
[\cos(\ell_E\cdot\tildek)+1]\\ 
&\times&\displaystyle 
\int_{4(\ell_E^2+m^2)}^{\infty}\frac{dk^{\prime 2}}{k^{\prime 2}}
\frac{2v^{\prime -1}}
{(k^{\prime 2}-k^2-i\epsilon)(k^{\prime 2}-\mu^2-i\epsilon)}\\
&=&\displaystyle 
(k^2-\mu^2)\frac{d}{dm^2}
\int_{4m^2}^{\infty}\frac{dk^{\prime 2}}{k^{\prime 2}}
\int\frac{d^n\ell_E}{(2\pi)^n}[\cos(\ell_E\cdot\tildek)+1]\\ 
&\times&\displaystyle 
v^{\prime -1}\tau(k^{\prime 2}-4(\ell_E^2+m^2))
\frac{2}{(k^{\prime 2}-k^2-i\epsilon)(k^{\prime 2}-\mu^2-i\epsilon)}, 
\end{array}
\end{equation}
with $v^{\prime}=\sqrt{1-4(\ell_E^2+m^2)/k^{\prime 2}}$. The above equation 
will be in the desired form if we can pass the operation $d/dm^2$ through the 
integral over $k^{\prime 2}$. We show below that this is indeed permitted since 
the resulting additional term corresponding to evaluating the integrand at 
$k^{\prime 2}=4m^2$ actually vanishes. To see this, consider the relevant 
radial factor of its $\ell_E$ integral at $k^{\prime 2}=4m^2(1+\eta)$ in the 
limit of $\eta\to 0^+$, 
\begin{equation}
\begin{array}{rcl}
{\rm integral}&\sim&\displaystyle 
\int d\ell_E \ell_E^{n-1}v^{\prime -1}\tau(k^{\prime 2}-4(\ell_E^2+m^2))\\
&\sim&\displaystyle 
\int_0^{\eta}dy y^{(n-2)/2}(\eta-y)^{-1/2}\\ 
&\propto&\displaystyle 
\eta^{(n-1)/2},
\end{array}
\end{equation}
which indeed vanishes as $\eta\to 0^+$. We have ignored the cosine factor which 
is smooth at $\ell_E\sim 0$. Actually, the vanishing of the additional term is 
just a reflection of the fact that ${\rm Im}~\Sigma$ vanishes at the physical 
threshold due to kinematical factors. The dispersion relation eq. 
$(\ref{selfenergy})$ is thus established at one loop in perturbation theory. 
We also notice that, since the subtraction is made with respect to $k^2$ but 
for a fixed value of $\tilde{k}^2$, the UV/IR mixing is automatically removed 
from the relation. 

\section{Dispersion relation for forward scattering} 

The aim of dispersion relations is to provide useful relations amongst 
measurable quantities on some basic assumptions and independently of 
perturbation theory in particular. It is already a difficult task to derive 
a dispersion relation for a general two-by-two scattering amplitude in 
axiomatic field theory on ordinary spacetime. 
While microcausality is a necessary requirement for this, it is 
usually not sufficient to guarantee simple dispersion relations. The amplitude 
depends on three Mandelstam variables which are constrained for a physical 
process. The analytic properties in one variable, which are the central part of 
dispersion relations, are inevitably entangled with those in other variables, 
making things involved. When one tries to represent dispersion integrals in 
terms of measurable quantities, one is usually afflicted with contributions 
from unphysical domains. In NC field theory the situation becomes more 
difficult and unclear. Even if we insist on other basic assumptions, we 
only have part of spacetime symmetries for special cases of 
$\theta_{\mu\nu}$, making the corresponding arguments less powerful. Related to 
this is the appearance of new variables involving NC momenta on which the 
amplitude can depend. These variables are 
not completely free parameters but are more or less constrained by ordinary 
variables for a physical process. This will cause new difficulties. 
Nevertheless, we would like to make a first attempt on this topic 
and show where the new difficulties may arise.

We consider the simplest case of forward scattering of a massless scalar 
particle against a massive one. This is the analogue of the forward 
photon-nucleon scattering in ordinary theory which has no unphysical regions in 
dispersion integrals. We assume that the massive particle is at rest in the 
reference frame where $\theta_{\mu\nu}$ is assigned its value. Note that this 
is a nontrivial assumption due to the loss of spacetime symmetries; we are only 
left with boosts in the direction perpendicular to the NC plane and rotations 
in the plane. And it is the simplest possibility in the sense that we only have 
one ordinary variable and one NC variable. Following the usual manipulation, 
the scattering amplitude is, 
\begin{equation}
\begin{array}{rcl}
{\cal A}&=&\displaystyle 
i\int d^4x~e^{ik\cdot x}\tau(x_0)\langle M|[j(x),j(0)]|M\rangle,
\end{array}
\end{equation}
up to terms which do not affect extraction of analytic properties in energy. 
$k^{\mu}=\omega(1,\vec{e}_k)$ with $\vec{e}_k\cdot\vec{e}_k=1$ is the 
four-momentum of the massless scalar with source $j$, and $M$ is the mass of 
the massive scalar. Inserting the completeness relation of the physical 
spectrum and using translational invariance, we have, 
\begin{equation}
\begin{array}{rcl}
\langle M|j(x)j(0)|M\rangle&=&\displaystyle 
\sum_{\alpha}\int\frac{d^3\vec{p}}{(2\pi)^32E_{\vec{p},\alpha}}
\langle M|j(x)|\vec{p},\alpha\rangle\langle\vec{p},\alpha|j(0)|M\rangle\\
&=&\displaystyle 
\sum_{\alpha}\int\frac{d^3\vec{p}}{(2\pi)^32E_{\vec{p},\alpha}}
|\langle M|j(0)|\vec{p},\alpha\rangle|^2e^{iMx_0}e^{-ip_+\cdot x}. 
\end{array}
\end{equation}
Denoting as $f_{\alpha}(\tildep^2)$ the matrix element squared which depends on 
$p$ only in the form of $\tildep^2$ according to our previous discussion, we 
obtain, 
\begin{equation}
\begin{array}{rcl}
\langle M|j(x)j(0)|M\rangle&=&\displaystyle 
\sum_{\alpha}e^{iMx_0}f_{\alpha}((i\tilde{\partial})^2)D(x,m^2_{\alpha}). 
\end{array}
\end{equation}
For $\theta_{0i}=0$, the derivatives do not involve $x_0$ so that 
\begin{equation}
\begin{array}{rcl}
{\cal A}&=&\displaystyle 
i\int d^4x~e^{ik\cdot x}\tau(x_0)\sum_{\alpha}
f_{\alpha}((i\tilde{\partial})^2)
\left[e^{iMx_0}D(x,m^2_{\alpha})-e^{-iMx_0}D(-x,m^2_{\alpha})\right], 
\end{array}
\end{equation}
Denoting the sum over $\alpha$ as 
$i^{-1}f(x_0,x_{\parallel}^2,\vec{x}^2_{\bot})$ which can 
depend on the relevant variables in the indicated way, we arrive at, 
\begin{equation}
\begin{array}{rcl}
{\cal A}&=&\displaystyle 
\int d^4x~e^{i\omega(x_0-\vec{e}_k\cdot\vec{x})}
\tau(x_0)f(x_0,x_{\parallel}^2,\vec{x}^2_{\bot}).
\end{array}
\label{eq_A}
\end{equation}

To proceed further, we have to impose some microcausality requirement to 
provide support properties in eq. $(\ref{eq_A})$ so that analytic 
continuation in energy may become possible. Since Lorentz invariance is 
generally lost on NC space with $\theta_{0i}=0$, it seems to make no sense 
to speak of space-like or time-like intervals when a nonzero interval occurs 
in the NC plane. As we described in the introduction, however, Lorentz 
invariance still remains in the normal direction to the NC plane. We thus 
propose the following microcausality requirement to replace the usual one, 
\begin{equation}
[j(x),j(0)]=0,~{\rm for}~x_0^2-x_3^2<0, 
\label{eq_mc}
\end{equation}
where we have assumed without loss of generality that NC is restricted to 
the $(12)$-plane. This is a weaker requirement than the usual one in the 
sense that a smaller region in spacetime is excluded as unphysical: while 
$x_0^2-x_3^2<0$ implies $x^2<0$, it makes no restriction on $x_{1,2}$. 

To show the viability of the above assumption, we compute the following 
commutator which arises at first order in perturbative $\varphi^3$ theory 
$\cite{chaichian}$, 
\begin{equation}
\begin{array}{rcl}
G&=&[(\varphi\star\varphi)(x),(\varphi\star\varphi)(y)], 
\end{array}
\end{equation}
where $\varphi$ is a free field with mass $m$. We have 
\begin{equation}
\begin{array}{rcl}
G&=&(\varphi(x)\varphi(y))\star i\Delta(z,m^2)
+i\Delta(z,m^2)\star(\varphi(y)\varphi(x))\\ 
&&+\varphi(x)\star_x i\Delta(z,m^2)\star_y\varphi(y)
+\varphi(y)\star_y i\Delta(z,m^2)\star_x\varphi(x), 
\end{array}
\end{equation}
where $z=x-y$, $\star=\star_x\star_y$ and $\star_{x(y)}$ refers to the star 
product with respect to $x(y)$. Since $\Delta(z,m^2)=0$ for $z^2<0$, it would 
be tempting to conclude that $G$ also vanishes. But this is actually very 
subtle due to the presence of the star product which involves derivatives 
of infinite order. The highly nonlocal character of the star product may 
very likely prohibit us from setting $z$ a space-like value before the 
multiplication is finished. It is not clear whether it is mathematically 
possible to circumvent this problem by a proper extension of distributions. 
Instead, what is clear is that $G$ vanishes for $z_0^2-z_3^2<0$: the star 
multiplication has nothing to do with $z_0$ and $z_3$ and is thus commutable 
with the procedure of setting a value to $z_0^2-z_3^2$, while 
$z_0^2-z_3^2<0$ guarantees $z^2<0$ and thus the vanishing of 
$\Delta$ and $G$. 

The above statement can be made more transparent. Let us consider the 
following typical quantity appearing in $G$, 
\begin{equation}
\begin{array}{rl}
&(g(x)h(y))\star D(z,m^2)\\
=&\displaystyle(2\pi)^{-3}\int d^2\vec{p}_{\bot}
\left(g(x)\star_x e^{+i\vec{p}_{\bot}\cdot\vec{x}_{\bot}}\right)
\left(h(y)\star_y e^{-i\vec{p}_{\bot}\cdot\vec{y}_{\bot}}\right)
\int\frac{dp_3}{2E_{\vec{p}}}e^{-i(E_{\vec{p}}z_0-p_3z_3)}, 
\end{array}
\end{equation}
where $g,h$ can be field operators or c-number functions. For $z_0^2<z_3^2$, 
we make a change of variable, 
$p_3=\gamma(p_3^{\prime}+\beta E_{\vec{p}^{\prime}})$ 
with $\beta=z_0/z_3$ and $\gamma=(1-\beta^2)^{-1/2}$, which amounts to a 
Lorentz boost of the momentum in the $3$-direction. Using 
$dp_3/E_{\vec{p}}=dp_3^{\prime}/E_{\vec{p}^{\prime}}$ 
and $E_{\vec{p}}z_0-p_3z_3=-p_3^{\prime}z_3^{\prime}$, with 
$z_3^{\prime}=\gamma^{-1}z_3$ and $z_0^{\prime}=0$, and dropping the prime 
in $\vec{p}$, the above becomes  
\begin{equation}
\begin{array}{l}
\displaystyle(2\pi)^{-3}\int d^2\vec{p}_{\bot}
\left(g(x)\star_x e^{+i\vec{p}_{\bot}\cdot\vec{x}_{\bot}}\right)
\left(h(y)\star_y e^{-i\vec{p}_{\bot}\cdot\vec{y}_{\bot}}\right)
\int\frac{dp_3}{2E_{\vec{p}}}e^{+ip_3z_3^{\prime}}. 
\end{array}
\end{equation}
For $(g(x)h(y))\star D(-z,m^2)$, upon $\vec{p}\to -\vec{p}$, it becomes equal 
to the above. We thus arrive at, 
\begin{equation}
(g(x)h(y))\star \Delta(z,m^2)=0, ~{\rm for}~z_0^2-z_3^2<0.
\end{equation}
This calculation also makes it clear that the commutator $G$ vanishes for 
any power of fields when $z_0^2-z_3^2<0$. This lends perturbative supports to 
the assumption in eq. $(\ref{eq_mc})$. 

Let us now return back to eq. $(\ref{eq_A})$. Due to the weaker 
microcausality, the integrand can be nonvanishing for $x_0>|x_3|$ 
($x_3=x_{\parallel}$). This is 
a larger region than the usual forward light cone, where the 
phase is neither positive- nor negative-definite. The direct analytic 
continuation to the complex $\omega$ plane is thus not possible. To make the 
phase positive-definite, we first finish integration over $\vec{x}_{\bot}$ and 
obtain formally, 
\begin{equation}
\begin{array}{rcl}
{\cal A}&=&\displaystyle 
\int dx_0dx_3~e^{i(\omega x_0-k_3x_3)}
\tau(x_0)g(x_0,x_3^2;\omega_{\bot}), 
\end{array}
\end{equation}
with $\omega_{\bot}=|\vec{k}_{\bot}|$. As the integrand is even in $x_3$, 
we may replace $k_3$ by its absolute value, 
\begin{equation}
\begin{array}{rcl}
{\cal A}(\omega;\omega_{\bot})&=&\displaystyle 
\int dx_0dx_3~e^{i(\omega x_0-\sqrt{\omega^2-\omega^2_{\bot}}x_3)}
\tau(x_0)g(x_0,x_3^2;\omega_{\bot}).
\end{array}
\end{equation}
The problem has been reduced to a $(1+1)$ dimensional one with an effective 
mass of $\omega_{\bot}$. Let us just press on to see whether it is possible 
to write down a dispersion relation for ${\cal A}$ as we deal with the 
forward scattering of massive particles in ordinary theory. Since 
${\rm Im}\sqrt{\omega^2-\omega^2_{\bot}}>{\rm Im}~\omega$ for 
${\rm Im}~\omega>0$, continuation to the upper half of the complex $\omega$ 
plane is still not possible for $\omega_{\bot}\ne 0$. To identify the 
problem, we write, 
\begin{equation}
\begin{array}{rcl}
{\cal A}(\omega;\omega_{\bot})&=&\displaystyle 
\int_0^{\infty} dr~{\cal A}(r,\omega;\omega_{\bot}), 
\end{array}
\end{equation}
where 
\begin{equation}
\begin{array}{rcl}
{\cal A}(r,\omega;\omega_{\bot})&=&\displaystyle 
2\cos\left(r\sqrt{\omega^2-\omega^2_{\bot}}\right)
\int_{r}^{\infty} dt~e^{i\omega t}~g(t,r^2;\omega_{\bot}).
\end{array}
\end{equation}

${\cal A}(r,\omega;\omega_{\bot})$ is analytic in $\omega$ for 
${\rm Im}~\omega>0$ and finite $r$, since the phase factor never blows up 
and the cosine factor is actually a function of 
$r^2(\omega^2-\omega^2_{\bot})$ which is analytic and does not introduce 
branch cuts. We can therefore write down a dispersion relation for it, 
\begin{equation}
\begin{array}{rcl}
{\cal A}(r,\omega;\omega_{\bot})&=&\displaystyle 
\frac{1}{\pi}\int_{-\infty}^{\infty}d\omega^{\prime}
\frac{{\rm Im}~{\cal A}(r,\omega^{\prime};\omega_{\bot})}
{\omega^{\prime}-\omega-i\epsilon}. 
\end{array}
\end{equation}
For ${\cal A}(r,\omega;\omega_{\bot})$ 
decaying not fast enough as $|\omega|\to\infty$, it is necessary to make 
subtractions with respect to $\omega$ and for fixed $r$ and 
$\omega_{\bot}$. To convert the above dispersion relation to the one 
for ${\cal A}(\omega;\omega_{\bot})$ we have to integrate over $r$. This 
would be done if it were permitted to interchange the order of $r$ and 
$\omega^{\prime}$ integration. Indeed, for 
$|\omega^{\prime}|>\omega_{\bot}$ this is permitted since the cosine factor 
remains finite in the whole region of $r$. However, this is not the case 
in the unphysical interval $|\omega^{\prime}|<\omega_{\bot}$ where the 
cosine factor blows up as $r\to\infty$. This is apparently similar to the 
situation with massive particles in ordinary theory up to an essential 
difference. Namely, while the problem is reduced to be $(1+1)$ dimensional, 
the mass shell conditions for physical states are still $(1+3)$ dimensional: 
the effective mass $\omega_{\bot}$ is not a fixed physical parameter but 
varies with the scattering configuration. This implies that the above 
unphysical interval could not be associated with a pole contribution as 
it is in ordinary theory. This difficulty resulting from a weaker 
microcausality prevents us from writting a dispersion relation for the 
scattering amplitude. 

Finally, we comment briefly on a special configuration and the case with a 
massive incident particle. For $\omega_{\bot}=0$ where the forward 
scattering occurs exclusively in the commutative direction, the above 
unphysical region disappears so that there is no difficulty to write a 
dispersion relation for the amplitude. This difference to ordinary theory, 
i.e., the dependence on the detailed kinematic configurations, can be 
traced to Lorentz symmetry violation in NC field theory. When the incident 
particle has the mass $m$, the above unphysical region splits into two 
subintervals. The contribution from the one with $|\omega^{\prime}|<m$ 
can still be attributed to a pole term as in ordinary theory with the 
modification that its residue depends on $\omega_{\bot}$. The subinterval 
with $m<|\omega^{\prime}|<\sqrt{\omega_{\bot}^2+m^2}$ is afflicted with 
the same problem discussed above and presents an obstacle to dispersion 
relations. 

\section{Conclusion and discussion} 

The phase-like nonlocal nature of NC field theory makes its analytic properties 
very different from ordinary theory. These properties are intimately 
related to the issues of unitarity and causality. In this work we tried to 
incorporate them in the spectral representation and dispersion relations by 
following the development in axiomatic field theory on ordinary spacetime. 

We analyzed spacetime symmetries on NC spacetime that provide a necessary piece 
of argument in deriving dispersion relations. The NC kinematic 
variables appear naturally as 
parametric ones and the analytic properties are studied with respect to the 
usual kinematic variables. This offers a close similarity to ordinary theory 
as is illustrated in our perturbative example: as required by unitarity, 
an imaginary part in amplitude develops only when a physical threshold is 
crossed. On the other hand, the symmetry analysis forces us to impose a 
microcausality requirement that is weaker than the usual one; namely, 
observables commute when they are separated by an interval that is space-like 
in the commutative direction. The viability of the assumption is supported 
by the perturbative calculation of a typical commutator. 
Starting from this weaker microcausality, however, there arises a new 
unphysical region when deriving dispersion relations for forward scattering. 
Since the region is specified by a kinematic variable in the NC plane instead 
of a mass parameter, it does not seem to be associated to a pole term. 
It therefore cannot be surmountable using the usual tricks in ordinary theory. 

Our treatment of ordinary variables and NC variables hints at a possible 
subtraction and renormalization scheme that may apply generally. Whenever 
possible, dispersion relations are written for fixed NC variables while 
ordinary variables are integrated over. If there is a UV divergence, 
subtractions are to be made at the same NC variables. The planar and 
nonplanar contributions are thus treated on the same footing as can be 
seen in our perturbative example. This is in contrast to the popular 
treatment in the literature that causes the UV/IR mixing problem. We also 
add in passing that the mixing is not necessarily associated with the 
infrared; if subtraction and renormalization is made with respect to a 
nonvanishing motion in the NC plane, the mixing enters when the motion is 
restricted in the commutative direction. This idea will be further 
elaborated elsewhere.


\begin{thebibliography}{30}

\bibitem{mixing}
S. Minwalla, M. V. Raamsdonk and N. Seiberg, 
{\it Noncommutative perturbative dynamics}, 
\JHEP 02 (2000) 020 [hep-th/9912072];
I. Ya. Aref'eva, D. M. Belov and A. S. Koshelev, 
{\it Two-loop diagrams in noncommutative $\phi_4^4$ theory}, 
\PL B476 (2000) 431 [hep-th/9912075]; 
M. V. Raamsdonk and N. Seiberg, 
{\it Comments on noncommutative perturbative dynamics}, 
\JHEP 03 (2000) 035 [hep-th/0002186];
A. Matusis, L. Susskind and N. Toumbas, 
{\it The UV/IR connection in the noncommutative gauge theories},
\JHEP 12 (2000) 002 [hep-th/0002075].

\bibitem{filk}T. Filk, {\it Divergences in a field theory on quantum space}, 
\PL B376 (1996) 53;
M. Chaichian, A. Demichev and P. Presnajder, 
{\it Quantum field theory on noncommutative space-times and the persistence 
of ultraviolet divergences}, 
\NP B567 (2000) 360 [hep-th/9812180].

\bibitem{unitarity}
J. Gomis and T. Mehen, 
{\it Space-time noncommutative field theories and unitarity}, 
\NP B591 (2000) 265 [hep-th/0005129]. 

\bibitem{bahns}D. Bahns, S. Doplicher, K. Fredenhagen and G. Piacitelli, 
{\it On the unitarity problem in space-time noncommutative theories}, 
\PL B533 (2002) 178 [hep-th/0201222]; see also: 
C. Rim and J. H. Yee, 
{\it Unitarity in space-time noncommutative field theories}, hep-th/0205193.

\bibitem{liao}Y. Liao and K. Sibold, 
{\it Time-ordered perturbation theory on noncommutative spacetime: 
basic rules}, hep-th/0205269; 
{\it Time-ordered perturbation theory on noncommutative spacetime II: 
unitarity}, hep-th/0206011; both to appear in \EPJ C. 

\bibitem{causality}N. Seiberg, L. Susskind and N. Toumbas, 
{\it Space-time noncommutativity and causality}, 
\JHEP 06 (2000) 044 [hep-th/0005015]; 
L. Alvarez-Gaume and J.L.F. Barbon, 
{\it Nonlinear vacuum phenomena in noncommutative QED}, 
\IJMP A16 (2001) 1123 [hep-th/0006209].

\bibitem{lubo}M. Lubo, 
{\it Maximally localized states and causality in noncommutative quantum 
theories}, \PR D65 (2002) 066003 [hep-th/0106018].

\bibitem{dispersion}For the development of dispersion relations 
and original papers, see for example: 
G. R. Screaton (Ed.), Dispersion relations (Oliver and Boyd, 1961); 
S. S. Schweber, An introduction to relativistic quantum field theory 
(Harper $\&$ Row, 1961);
J. D. Bjorken and S. D. Drell, Relativistic quantum fields 
(McGraw-Hill, 1965). 

\bibitem{brandt}F. T. Brandt, A. Das and J. Frenkel, 
{\it Dispersion relations for the self-energy in noncommutatie field theories}, 
hep-th/0206058.

\bibitem{chaichian}For a similar but different calculation of the commutator 
and its implications, see: M. Chaichian, K. Nishijima and A. Tureanu, 
{\it Spin statistics and CPT theorems in noncommutative field theory}, 
hep-th/0209008.
\end{thebibliography}
\end{document}